\pdfoutput=1
\documentclass[
    reprint,
    aps,
    preprintnumbers,
    twocolumn,
    prb,
    superscriptaddress
]{revtex4-2}

\usepackage[utf8]{inputenc}
\usepackage{booktabs}
\usepackage[multiple]{footmisc}
\usepackage{rotating}
\usepackage{perpage}
\usepackage{amssymb}
\usepackage{amsbsy}
\usepackage{amsmath}
\usepackage{tikz}
\usepackage[T1]{fontenc}
\usepackage{etoolbox}
\usepackage{graphics}
\usepackage{abraces}
\usepackage[
  separate-uncertainty=true,
  per-mode=symbol-or-fraction,
]{siunitx}
\DeclareSIUnit\angstrom{Å}
\usepackage{braket}
\usepackage{hyperref}
\usepackage{multirow}
\usepackage{mathtools}
\usepackage{bm}
\usepackage{url}
\usepackage{physics}
\usepackage{hyperref}
\usepackage{bbm}
\usepackage{color}
\usepackage{placeins}
\usepackage[normalem]{ulem}

\newcommand{\mat}[1]{\boldsymbol{#1}}
\renewcommand{\vec}[1]{\boldsymbol{#1}}
\AtBeginDocument{\renewcommand{\Re}{\operatorname{Re}}}
\newcommand{\dx}[1]{\,\text{d}#1}
\newcommand{\dl}{_\text{del}}

\begin{document}

\title{Optical Pumping of Bardeen-Cooper-Schrieffer Superconductors}


\author{Vanessa Sulaiman}\email{vanessa.sulaiman@tu-dortmund.de}
\affiliation{Condensed Matter Theory, TU Dortmund University,
Otto-Hahn Stra\ss{}e 4, 44227 Dortmund, Germany}

\author{G\"otz S.~Uhrig}
\email{goetz.uhrig@tu-dortmund.de}
\affiliation{Condensed Matter Theory, TU Dortmund University,
Otto-Hahn Stra\ss{}e 4, 44227 Dortmund, Germany}

\date{\today}


\begin{abstract}
  Motivated by the generation by optical pulses of non-thermal distributions of nuclear spins in quantum dots
  we investigate the effect of optical pulses applied to Bardeen-Cooper-Schrieffer (BCS) superconductors.
  Using time-dependent mean-field theory formulated with Anderson pseudospins,
  we study the electronic configurations and the energy deposited in the system by optical pulses.
  The pulses are included by Peierls substitution and we study short
  rectangular pulses as well as idealized $\delta$ pulses.
	Already a few and even a single pulse generates highly non-trivial distributions of electron expectation values
  which we simulate numerically and explain analytically based on the linearization of the equations of motion.
  These results suggest so far unexplored experimental possibilities for the optical control of superconducting states.
\end{abstract}

\maketitle


\section{Introduction}\label{sec:Intro}

Since the Bardeen-Cooper-Schrieffer (BCS) theory of superconductivity was first established \cite{bardeen1957},
many different experiments with continuous irradiation \cite{dayem1967, dahlberg1979, parker1972} were carried out
to investigate whether an increase of the critical current and of the energy gap can be reached in this way.
In thin-film bridges, an increase of the critical current could be reached \cite{dayem1967, dahlberg1979}.
However, this approach did not enhance the gap in a measurable way in similar experiments with tunnel junctions
\cite{dahlberg1979}.

Recent advances in the generation of ultrashort pulses have enabled further studies using terahertz (THz) pump-probe setups
\cite{beyer2011, beck2013}.
These pulses allow the investigation of non-adiabatic excitations, for instance, for
exciting Higgs modes, because they are fast enough to shake the Cooper pair condensate and their THz carrier frequency
is in the same range as the typical gap energy ${\Delta/h}$ divided by the Planck constant $h$.
This has renewed the theoretical interest in the BCS model, in particular in the non-equilibrium regime.
Many studies investigated quenches, i.e., instantaneous changes of the order parameter or the coupling constant.
Some focus on the resulting dynamics of the order parameter \cite{yuzbashyan2005, dzero2009, kirmani2019, cui2019, cui2019a},
others on the excitation of collective modes such as the Higgs mode or the
Anderson-Bogoliubov (phase) mode \cite{tsuji2015, muller2019, derendorf2022}.
In addition, many studies on the response induced by external electromagnetic fields were carried out
\cite{gangopadhyay2010, li2024, derendorf2024}.

The mean-field dynamics of the BCS model share great similarities with those of the central spin model \cite{yuzbashyan2005a}.
The latter describes the electron spin dynamics in singly charged quantum dots for which
it has been established experimentally \cite{greil06b,greil07a,greil07b}
and theoretically \cite{beuge16,jaschke2017, beuge17, kleinjohann2018, schering2020}
that periodic pulses lead to quasi-stationary states  favoring periodic revival of the spin polarization.
This enhances the coherence of the central spin significantly which can be useful for applications
such as quantum state purification \cite{uhrig2020}.
The appearance of commensurate structures in the distribution of nuclear spins of quantum dots by optical pulses
motivates us to look for similar non-trivial effects in related systems.
Here we investigate which structures are induced by optical pulses applied to BCS superconductors.
Hence, this study is complementary to the numerous studies investigating THz pulses.

We investigate the effect of optical pulses on the electronic configurations in the BCS model.
Two different pulse shapes are considered: a rectangular pulse and an idealized $\delta$ pulse.
The aim is to simulate and to analyze the pulse-induced distribution of expectation values theoretically
in order to motivate experimental studies of this type and to predict the results to be expected.
This will extend the possibilities of actively controlling matter.

This article is organized as follows:
In Sect.~\ref{sec:Model}, we introduce the model including the description of the electromagnetic pulse.
Subsequently, we investigate rectangular pulses in Sect.~\ref{sec:RectangularPulse} and in particular
the influence of the pulse duration.
In Sect.~\ref{sec:DeltaPulse}, we examine an instantaneous $\delta$ pulses and the influence of the delay time.
We conclude our analysis in Sect.~\ref{sec:Conclusion}.


\section{Model}
\label{sec:Model}

The Hamiltonian for $s$-wave BCS superconductors is given by
\begin{equation}
    \label{eq:BCS}
    H_\text{BCS} = \sum_{k,\sigma}\varepsilon_k c_{k,\sigma}^\dagger c_{k,\sigma}
    - \frac{V}{N}\sum_{k,k'} c_{k,\uparrow}^\dagger c_{-k,\downarrow}^\dagger c_{-k',\downarrow}c_{k',\uparrow} ,
\end{equation}
where $c_{k,\sigma}^{(\dagger)}$ are the fermionic annihilaton (creation) operators
with spin ${\sigma \in \{\uparrow,\downarrow\}}$
and $\varepsilon_k$ the electronic dispersion.
The second term describes the attractive electron-electron interaction with $V > 0$ between the electrons of the Cooper pairs.
It results from the elimination of the electron-phonon interaction, for instance by unitary
transformations, see, e.g., Ref.~\onlinecite{krull2012}.
Since the relevant energies are much smaller than the Fermi energy $\varepsilon_{\text{F}}$,
we consider the energy difference
\begin{equation}
    \varepsilon_k = \frac{\hbar^2k^2}{2m} - \varepsilon_\text{F} .
\end{equation}
The order parameter
\begin{equation}
    \Delta = \frac{V}{N}\sum_{k'} \langle c_{-k',\downarrow} c_{k',\uparrow} \rangle
\end{equation}
is also the energy gap of the system.
The binding energy of a single Cooper pair is given by $2\Delta$ which is also the energy of a Higgs boson \cite{pekke15,krull16}.

The Anderson pseudospin operators~\cite{anderson1958,yuzbashyan2005a}
\begin{subequations}
    \begin{align}
        \sigma_k^z &= \left(c_{k,\uparrow}^\dagger c_{k,\uparrow} + c_{-k,\downarrow}^\dagger c_{-k,\downarrow} - 1\right)/2 \\
        \sigma_k^+ &= c_{k,\uparrow}^\dagger c_{-k\downarrow}^\dagger \\
        \sigma_k^- &= c_{-k,\downarrow} c_{k,\uparrow}
    \end{align}
\end{subequations}
have all properties of spin-$\frac{1}{2}$ operators.
Thus, we can utilize the same methods as for spin systems to analyze the Hamiltonian and its dynamics.
We rewrite the Hamiltonian with the pseudospin operators in the form
\begin{equation}
    H_\text{BCS} = \sum_k 2\varepsilon_k\sigma_k^z - \frac{V}{N} L_+ L_- \quad\text{with}\quad \vec{L} = \sum_k \vec{\sigma}_k
\end{equation}
where we used bold face symbols to denote vectors which we continue to do below.
Then, the self-consistency implies
\begin{equation}
  \Delta(t) = \Delta_x(t) - i\Delta_y(t) = \frac{V}{N}\sum_{k'} \langle \sigma_k^-(t) \rangle .
\end{equation}

Now we apply mean-field BCS theory where the effective field seen by each pseudospin $\vec{\sigma}_k$ is replaced by
the quantum mechanical expectation values
\begin{equation}
  \vec{b}_k(t) = (-2\Delta_x(t), -2\Delta_y(t), 2\varepsilon_k)^\top  .
\end{equation}
The resulting Hamiltonian
\begin{equation}
  H_\text{BCS} = \sum_k\vec{b}_k\cdot\vec{\sigma}_k
  \label{eq:BCS_MF}
\end{equation}
induces the equations of motion
\begin{equation}
  \dot{\vec{\sigma}}_k = i[H_\text{BCS}, \vec{\sigma}_k] = \vec{b}_k\times\vec{\sigma}_k  .
\end{equation}
Now, we can easily take the quantum mechanical expectation value
\begin{equation}
  \vec{s}_k(t) = 2\langle \vec{\sigma}_k(t)\rangle
  \label{eq:def_s}
\end{equation}
because the equations of motion are linear.
The resulting classical equations read
\begin{equation}
  \dot{\vec{s}}_k = \vec{b}_k \times \vec{s}_k
\end{equation}
with
\begin{align}
  \vec{b}_k &= (-VJ_x, -VJ_y, 2\varepsilon_k)^\top
  \intertext{and}
  \vec{J} &= \frac{1}{N}\sum_{k'}\vec{s}_{k'}  .
\end{align}

As the complex phase of the order parameter is constant due to particle-hole symmetry~\cite{barankov2006},
we can choose $\Delta$ such that $\Delta$ is real, i.e., ${\Delta_y = 0}$, for simplicity.
Then the self-consistency equation simplifies to
\begin{equation}
\label{eq:self-con2}
  \Delta(t) = \frac{V}{N}\sum_{k'} \frac{s_k^-(t)}{2} = \frac{1}{2}\frac{V}{N}\sum_{k'}s_{k'}^x (t)  .
\end{equation}
This also implies ${J_y = 0}$ and the equations of motion reduce to
\begin{subequations}\label{eq:dgl}
  \begin{align}
    \dot{s}_k^x &= -2\varepsilon_k s_k^y \\
    \dot{s}_k^y &= 2\varepsilon_k s_k^x + V J_x s_k^z\\
    \dot{s}_k^z &= -V J_x s_k^y  .
  \end{align}
\end{subequations}

The initial conditions are given by the ground state of the BCS model
\begin{equation}
  \langle c_{k,\uparrow}^\dagger c_{k,\uparrow} \rangle = \frac{1}{2} -\frac{\varepsilon_k}{2E_k}, \quad
  \langle c_{-k,\downarrow}c_{k,\uparrow} \rangle = \frac{\Delta}{2E_k}
\end{equation}
with the quasiparticle energy $E_k = \sqrt{\varepsilon_k^2 + \lvert\Delta\rvert^2}$.
The equilibrium state is characterized by
\begin{equation}
  s_k^x = \frac{\Delta}{E_k} , \quad s_k^y = 0, \quad s_k^z = -\frac{\varepsilon_k}{E_k}
  \label{eq:initial_config}
\end{equation}
for the Anderson pseudospins.
Note that these values represent a static solution of the self-consistency equations \eqref{eq:self-con2}.


We use the Peierls substitution~\cite{peierls1933} to include the electromagnetic field of the pulse in the model.
First, we consider the tight-binding model
\begin{equation}
  H_0 = - \sum_{i,j = 1}^N t_{ij}c_{i}^\dagger c_{j} - \mu \sum_{i = 1}^N c_i^\dagger c_i  .
\end{equation}
The electromagnetic field couples the hopping to $t_{ij}$.
Using the Hamiltonian gauge~\cite{jackson2001} we set ${\phi(\vec{r}, t) = 0}$
and the hopping term is changed by phase resulting from the vector potential.
This vector potential for a sinusoidal field is given by
\begin{equation}
  \vec{A}(t) = \frac{cE}{\gamma} \cos(\gamma t)  ,
\end{equation}
where $\gamma$ is the angular frequency and $E$ the amplitude of this electromagnetic field
while the speed of light is given by $c$.
For calculational simplicity, we consider a diagonal orientation of the vector potential,
i.e., $\vec{A}$ is oriented in the (1,1,1) spatial direction
so that all nearest-neighbor hopping elements are affected in the same way.
Inserting the vector potential into the Peierls substitution
\begin{equation}
  t_{ij} \to t_{ij} \exp\left( -\frac{ie}{\hbar c} \int_{\vec{R}_i}^{\vec{R}_j} \vec{A}(\vec{r},t) \dx{\vec{r}} \right)
\end{equation}
and averaging over the oscillations~\cite{kalthoff2018} in the spirit of a Magnus expansion \cite{blane09}, we obtain
\begin{equation}
  t_{ij} \to t_{ij} J_0\left(\frac{E_0}{\gamma}\right) \quad \text{with} \quad E_0 = \frac{eaE}{\hbar}
\end{equation}
with the zeroth Bessel function $J_0$ and the lattice constant $a$.

Averaging over the oscillations is well justified because we focus on optical pulses.
Even though they are short on an absolute scale of picoseconds, they are still long
relative to a single oscillation period and, expressed reversely, the oscillations are fast
relative to the pulse duration.
In addition, they are fast relative to the internal time scales of the superconductors.
Their energy (${\hbar\gamma\approx \SI{1}{eV}}$) is by about
three orders of magnitude higher than the value given by the superconducting gap ${\Delta \approx \SI{1}{meV}}$.
The corrections to the averaged Hamiltonian can be estimated to be of the order ${\Delta/(\hbar \gamma)}$.

In the BCS model, the effect of the optical pulse on the energies $\varepsilon_k$ consists in the
reduction~\cite{peierls1933,kalthoff2018}
\begin{equation}
  \varepsilon_k \to \varepsilon_k J_0\left(\frac{E_0}{\gamma}\right)
\end{equation}
during the pulse.
Our estimate is that the argument ${E_0 / \gamma}$ of the Bessel function is much less than 1 so that $J_0$ is close to 1.


In experiments, a typical material~\cite{matsunaga2012} is NbN with a lattice constant of
${a_\text{NbN} = \SI{4.39e-10}{\meter}}$.
Usually, an optical pulse~\cite{bossini2014} has a photon energy of ${\hbar\gamma = \SI{2.2}{eV}}$ and a
length of ${t_\text{puls} = \SI{100}{fs}}$ or longer.
The amplitude is approximately ${E \approx \SI{9.5e8}{\volt\per\meter}}$.
With these parameters, we obtain approximately ${\gamma t_\text{puls}/(2\pi) \approx 50}$ oscillations during one pulse.
Thus, we consider the approximation to use the averaged Hamiltonian as well justified for pulses of at least ${t_\text{puls} = \SI{100}{fs}}$.
If the pulse duration reaches $\SI{1}{ps}$ it comprises even 500 oscillations.
In addition, we estimate the argument of the Bessel function
\begin{equation}
    \frac{eaE}{\hbar\gamma} \approx \num{0.19} ,
\end{equation}
yielding
\begin{equation}
    J_0 \approx \num{0.99} .
\end{equation}


\section{Rectangular pulse}
\label{sec:RectangularPulse}

\begin{figure}[htb]
  \centering
  \includegraphics[width=\columnwidth]{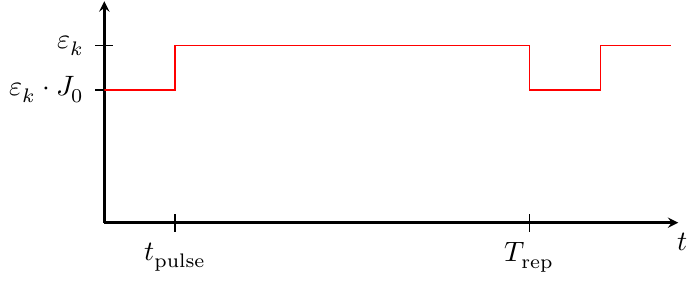}
  \caption{Time dependence of the effective $\varepsilon_k$ due to the Peierls substitution
	and the averaging for a rectangular pulse.}
  \label{fig:rectangle}
\end{figure}

We consider a rectangular pulse as shown in Fig.~\ref{fig:rectangle}.
Using the Peierls substitution, the energies $\varepsilon_k$ are modified during the pulse by the
constant factor $J_0$.
As we consider a diagonal orientation of the vector field,
all components of the pseudospin are affected by this substitution which simplifies the calculation.
We do not expect that this particular orientation has a big effect on the results.

Unless otherwise stated, the initial value of the order parameter $\Delta_0$ is chosen to be $\SI{1}{meV}$.
The BCS interaction is constant up to a cutoff that is given by the Debye frequency \cite{krull2012}
which we choose to be ${\varepsilon_\text{c} = \SI{50}{meV}}$ for the dispersion.
The simulation was realized for ${N = 10^5}$ equidistant discrete energies between
${-\varepsilon_\text{c}}$ to $\varepsilon_\text{c}$ measured relative to the Fermi energy.
The results will illustrate that it is indeed necessary to choose such a fine discretization.

\begin{figure}[htb]
    \centering
    \includegraphics[width=\columnwidth]{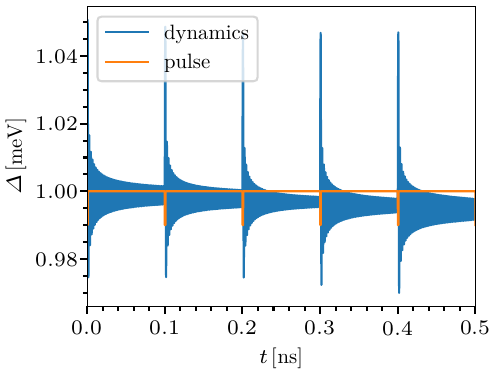}
    \caption{Dynamics of the superconducting order parameter.
            The pulse itself is shown in orange in arbitrary units.
            The blue oscillations are too fast to be resolved.
            The initial gap is  ${\Delta_0 = \SI{1}{meV}}$, the pulse depth is ${J_0 = \num{0.99}}$
            and the pulse duration ${t_\text{puls} = \SI{1}{ps}}$.}
    \label{fig:ex_dynamics}
\end{figure}

To illustrate the effect of a train of rectangular pulses, we choose the value of the Bessel function
to be ${J_0 = 0.99}$, the pulse duration  ${t_\text{puls} = \SI{1}{ps}}$, and
the repetition rate of the pulses ${T_\text{rep} = \SI{100}{ps}}$.
The resulting dynamics of the order parameter $\Delta$ are shown in Fig.~\ref{fig:ex_dynamics}.
It consists of fast sinusoidal oscillations which are not resolved in Fig.~\ref{fig:ex_dynamics} with an
envelope decaying like $\frac{1}{\sqrt{t}}$ similar to the dynamics after a quench
\cite{yuzbashyan2006, barankov2006, muller2018, akbari2013, kirmani2019, cui2019}.
The decay differs slightly from pulse to pulse and
the average order parameter $\Delta_\text{ave}$ around which the order parameter oscillates decreases with each pulse.
This is an expected behavior because quenches yield the long-time limit
${\Delta_\infty < \Delta_0}$~\cite{yuzbashyan2006, cui2019}.
So it is natural that each pulse reduces the order parameter by a certain amount
because it perturbs the superconducting order and deposits energy in the system, see also below.

\begin{figure}[htb]
    \centering
    \includegraphics[width=\columnwidth]{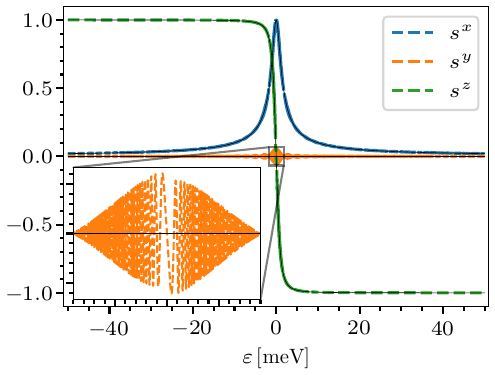}
    \caption{Expectation values of the pseudospins at ${t=0}$ (thin black solid lines)
            and at ${t=T_\text{rep}}$, i.e.,
						just before the second pulse (thick dashed colored lines).}
    \label{fig:ex_sy1}
\end{figure}

\begin{figure}[htb]
    \centering
    \includegraphics[width=\columnwidth]{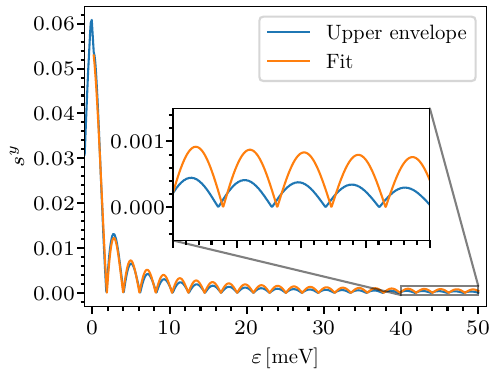}
    \caption{Upper envelope of the $y$ component of the spin distribution after the first pulse
            and a delay time ${t_\text{del}=T_\text{rep}}$. It is fitted by a sinc function as given
						in Eq.\ \eqref{eq:sinc}.}
    \label{fig:sinc_fit_example}
\end{figure}

In Fig.~\ref{fig:ex_sy1} the distribution of the expectation values of the pseudospin just before the second pulse,
i.e., at ${t=T_\text{rep}}$, is depicted.
For comparison, the initial distribution~\eqref{eq:initial_config} is shown in black.
After the pulse, some pseudospin values are amplified while others are suppressed depending on $\varepsilon_k$.
This can be seen in more detail in Fig.~\ref{fig:sinc_fit_example}.
The distances between the peaks seem to be constant while the height of the peaks
decreases with increasing energy $\varepsilon_k$.
In addition, there is a finer structure in the peaks that is shown in the inset.
To examine this further, we will focus on $s_y$ because the effects are easier to discern
in the $y$ component than in the other components because the $y$ component is
initially zero for all $\varepsilon_k$.

\subsection{Pseudospin distribution}
\label{sub:Spin configuration}

To analyze the pseudospin distribution after the first pulse, we examine its envelope first.
The behavior strongly resembles a sinc function
\begin{equation}
  f(x) = a \left\lvert \frac{\sin(b(x+c))}{x+c} \right\rvert
  \label{eq:sinc}
\end{equation}
so that we use it as fit function.
The resemblance of the envelope to a sinc function provides evidence
that the envelope is essentially given by the Fourier transform of the rectangular pulse shape.
This can indeed be underpinned  by an analytical approximation for the pseudospin distribution
the details of which are provided in App.~\ref{sec:A1}.
We find  that the $y$ component after a pulse can be approximated by
\begin{equation}
  \begin{split}
  s_{k,1}^y(t)
    &= 2 \frac{\Delta_0}{E_{k,0}} (J_0 - 1) \varepsilon_{k,0} t_\text{puls} \\
     &\hphantom{=}\cdot \cos\left( 2E_{k,0} \left(t - \frac{t_\text{puls}}{2}\right) \right)
		\mathrm{sinc}\left( E_{k,0} t_\text{puls} \right) .
  \end{split}
  \label{eq:sinc_analytical}
\end{equation}
The fit as displayed in Fig.~\ref{fig:sinc_fit_example}
is done for negative and positive energies $\varepsilon_k$ separately.
It is noticeable that the envelope decays slightly faster than the sinc function.
For low energies, the fit remains slightly below the envelope, but for higher energies,
the amplitude of the fit is above the envelope.
In contrast, the positions of the nodes are quite accurately reproduced by the fit.

\begin{figure}[htb]
    \centering
    \includegraphics[width=\columnwidth]{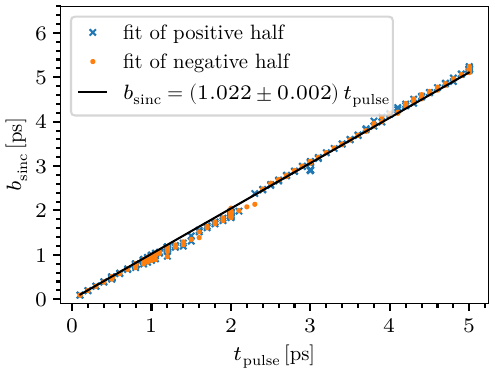}
    \caption{Fit parameter $b_\text{sinc}$ found from the sinc fit plotted as function of the pulse duration $t_\text{puls}$.
    The linear fit shows the agreement between both quantities.}
    \label{fig:sinc_fit_param}
\end{figure}

The finding that the sinc function in Eqs.~\eqref{eq:sinc} and \eqref{eq:sinc_analytical}, respectively,
results from the Fourier transform indicates that the distance $\Delta\varepsilon$ between the nodes depends on the pulse duration $t_\text{puls}$.
Hence, we expect
\begin{equation}
  \Delta \varepsilon = \frac{\pi}{b}  = \frac{\pi}{t_\text{puls}}.
  \label{eq:delta_eps}
\end{equation}
This relation is confirmed by the fit parameters for simulations with various $J_0$ and
pulse durations $t_\text{puls}$ in Fig.~\ref{fig:sinc_fit_param} where a very
good agreement between the fitted $b_\text{sinc}$ and $t_\text{puls}$ is established.

\subsection{Energy deposited by a pulse}
\label{sub:Energy difference}

In addition to the distribution of the pseudospin values, we are interested in the energy deposited by the pulse in the superconductor.
To this end, we must know the total energy for given pseudospin values.
In mean-field theory, the mean-field Hamiltonian in Eq.~\ref{eq:BCS_MF} reads
\begin{equation*}
    H_\text{MF} = \sum_k \vec{b}_k\,\cdot\vec{\sigma}_k \quad\text{with}\quad \vec{b}_k
		= \begin{pmatrix}-2\Delta_x \\ -2\Delta_y \\ 2\varepsilon_k\end{pmatrix} .
\end{equation*}
Taking its expectation value~\eqref{eq:def_s} leads to
\begin{equation}
    E = \langle H \rangle = \frac{1}{2} \sum_k \tilde{\vec{b}}_k\,\cdot\vec{s}_k  .
\end{equation}
But this overcounts the interaction energy by a factor of two \cite{gross1991}.
To obtain the correct energy, $\vec{b}_k$ has to be modified according to
\begin{equation}
    \tilde{\vec{b}}_k = (-\Delta_x,\, -\Delta_y,\, 2\varepsilon_k).
\end{equation}
Using the initial value ${\Delta_y = 0}$ and
\begin{equation}
    \Delta_x = \frac{V}{2N} \sum_{k'}s_{k'}^x = \frac{V}{2} J_x
\end{equation}
we can simplify $E$ to
\begin{equation}
    E = \sum_k \left(\varepsilon_k s_k^z-\frac{V}{4} J_x s_k^x\right) .
\end{equation}
With this relation, we evaluate the energy per pseudospin before and after the pulse
to determine the energy differences and sum over all of them.

\begin{figure}[htb]
    \centering
    \includegraphics[width=\columnwidth]{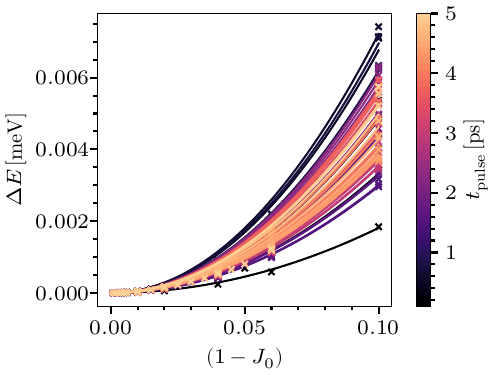}
    \caption{Energy difference induced by one pulse as a function of the pulse depth ${(1-J_0)}$.
            Each color represents a different pulse duration $t_\text{puls}$.
            Note the non-monotonic evolution of the parabolae upon increasing pulse duration.}
    \label{fig:energy_J0}
\end{figure}

In Fig.~\ref{fig:energy_J0}, the total energy difference is plotted as a function of the
pulse depth $J_0$ for various pulse durations $t_\text{puls}$ encoded by the given color scale.
The lines are quadratic fits
\begin{equation}
\label{eq:quadratic_fit}
    \Delta E(1-J_0) = a (1-J_0)^2
\end{equation}
which describe the dependence nicely.
This is in accordance with Fermi's golden rule which states that the energy intake is proportional to the modulus of the squared matrix element.
The latter is proportional to the pulse depth $(1-J_0)$.
Close inspection of Fig.~\ref{fig:energy_J0} reveals that there is no monotonic behavior
of the prefactor $a$ as function of the pulse duration $t_\text{puls}$.
This observation requires further investigation.

\begin{figure}[htb]
    \centering
    \includegraphics[width=\columnwidth]{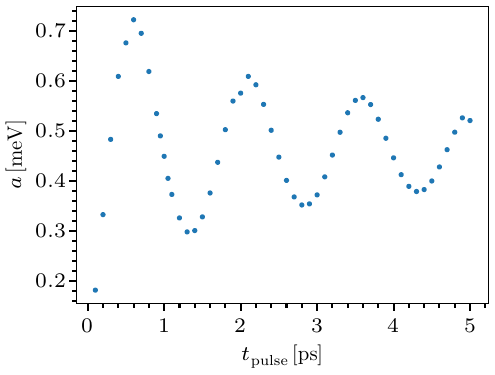}
    \caption{Fit parameter $a$ of the quadratic fit \eqref{eq:quadratic_fit} of Fig.~\ref{fig:energy_J0}.
    The non-monotonic oscillatory behavior is clearly visible.}
    \label{fig:energy_J0_params}
\end{figure}

\begin{figure}[htb]
    \centering
    \includegraphics[width=\columnwidth]{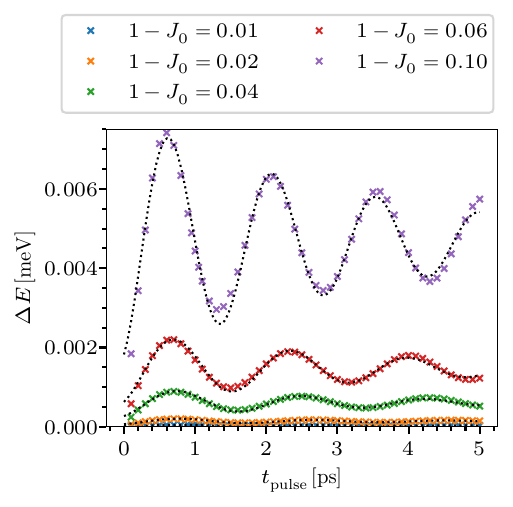}
    \caption{Energy difference induced by one pulse as a function of the pulse duration $t_\text{puls}$.
      The black dotted lines show the fits to the data according to Eq.~\ref{eq:energy-fit}.}
    \label{fig:energy_dt}
\end{figure}

In Fig.~\ref{fig:energy_J0_params} the parameter of the fits of Fig.~\ref{fig:energy_J0} are plotted
against the pulse duration.
There is a clear oscillatory dependence on the pulse duration  $t_\text{puls}$ discernible.
The same oscillations can be seen in the explicit energy difference as a function of
the pulse duration $t_\text{puls}$ for given pulse depth as is depicted in Fig.~\ref{fig:energy_dt}.
Here, we also find that the frequency of the oscillation does not depend significantly on the pulse depth $J_0$.
To quantify this behavior, we fit the data by the function
\begin{equation}
  \Delta E( t_\text{puls} ) = a\sin(b t_\text{puls} + \varphi) e^{-d t_\text{puls}} + c
  \label{eq:energy-fit}
\end{equation}
in order to extract the angular frequencies $b$; they are given in Tab.~\ref{tab:energy}.
The left table illustrates the weak dependence on the pulse depth.
The right table shows that the oscillation frequency scales with the initial order parameter $\Delta_0$.
There is roughly a factor of 2 between both quantities ${\hbar b/\Delta_0 \approx 2}$.
Our data (not shown) indicates that this factor applies to weak pulses and grows
upon increasing pulse depth.

\begin{table}[htb]
    \centering
    \begin{tabular}{S[table-format=1.2] S[table-format=1.2] @{${}\pm{}$} S[table-format=1.2]}
      \toprule
      {$1-J_0$} & \multicolumn{2}{c}{$b \, [\si{meV}\mathbin{/}\hbar]$} \\
      \midrule
      0.01 & 2.14 & 0.02 \\
      0.02 & 2.21 & 0.02 \\
      0.04 & 2.37 & 0.01 \\
      0.06 & 2.52 & 0.01 \\
      0.10 & 2.84 & 0.01 \\
      \bottomrule
    \end{tabular}
    \hspace{0.1em}
    \begin{tabular}{S[table-format=1.2] S[table-format=2.2] @{${}\pm{}$} S[table-format=1.2] S[table-format=2.2] @{${}\pm{}$} S[table-format=1.2]}
      \toprule
      {$\Delta_0 \, [\si{meV}]$} & \multicolumn{4}{c}{$b \, [\si{meV}\mathbin{/}\hbar]$} \\
      {} & \multicolumn{2}{c}{${1-J_0}=0.01$} & \multicolumn{2}{c}{${1-J_0}=0.04$} \\
      \midrule
      0.25 & 0.5  & 0.1  & 0.55 & 0.01 \\
      0.5  & 1.10 & 0.01 & 1.25 & 0.01 \\
      1.0  & 2.14 & 0.02 & 2.37 & 0.01 \\
      2.0  & 4.17 & 0.02 & 4.53 & 0.02 \\
      4.0  & 8.20 & 0.02 & 8.74 & 0.02 \\
      \bottomrule
    \end{tabular}
    \caption{Angular frequency $b$ of the deposited energy as a function of the pulse depth ${1-J_0}$ (left table)
		        for initial ${\Delta_0=\SI{1}{meV}}$ and as a function of the initial order parameter $\Delta_0$ (right table)
						for pulse depths ${{1-J_0}=0.01}$ and ${{1-J_0}=0.04}$.
						}
    \label{tab:energy}
\end{table}

\begin{figure}[htb]
    \centering
    \includegraphics[width=\columnwidth]{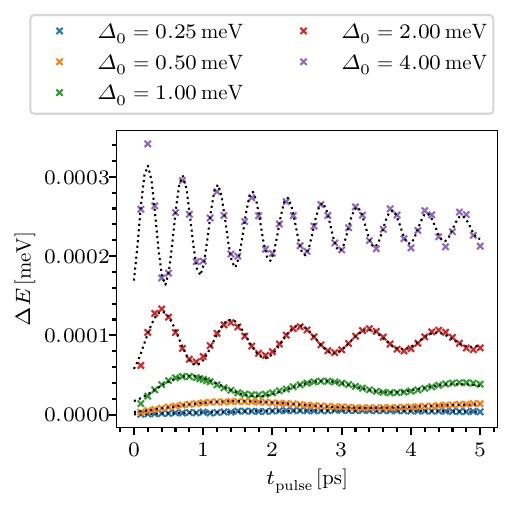}
    \caption{Energy difference induced by one pulse as a function of the pulse duration $t_\text{puls}$
            for various initial order parameters $\Delta_0$ at a pulse depth of ${1-J_0 = \num{0.01}}$.
            The black dotted lines indicate the fits according to Eq.~\ref{eq:energy-fit}.}
    \label{fig:energy_compare_delta0}
\end{figure}

For completeness, we provide results for the deposited energy for various
initial values of $\Delta_0$ in Fig.~\ref{fig:energy_compare_delta0} vs.\  the pulse duration.

These observations clearly reveal that the pulses induce to a precession of the pseudospins
which depends on the duration of the pulse and on the initial order parameters.
Naively, one would have assumed that longer pulses always change the system more than shorter pulses,
i.e., put more energy into them.
This is suggested by the naive application of Fermi's golden rule neglecting coherent effects.
Yet, this turns out to be not appropriate.
The dynamics we are investigating here is coherent on the time scale of the pulses and the considered delay.
Note that we have not included a relaxation mechanism in the model beyond dephasing.
Thus, the pulse can put energy into the superconductor, but it can also take it out again.
We find that the angular frequency of these oscillations corresponds roughly to
${b \approx 2\Delta_0/\hbar}$ which is the frequency of the Higgs mode.
The ratio ${\hbar b/\Delta_0}$ is closest to 2 for weak pulses with a decreasing trend for increasing gap values,
see Tab.~\ref{tab:energy}.


\section{\texorpdfstring{$\delta$ pulse}{𝛿 pulse}}
\label{sec:DeltaPulse}

Until now, we varied the parameters of the pulse, but
not the repetition time of the pulses $T_\text{rep}$ between the pulses.
In order to avoid an overwhelming amount of structural details we
want to simplify the pulses before addressing the influence of $T_\text{rep}$.
We achieve this by eliminating the dependence on the pulse duration by using
idealized $\delta$ pulses, i.e., instantaneous pulses.
We approximate the rectangular pulse by a $\delta$ pulse with the same area ${(1-J_0) t_\text{puls}}$.
Note that this appears counterintuitive since an infinite depth ${1-J_0\to \infty}$ does
not make sense as renormalization of the dispersion by an optical pulse.
But we will show that this idealization of rectangular pulses works quite well
and helps simplifying further investigations.

At first, we rewrite the time-dependent $\varepsilon_k$ as
a sum of a constant and a time-dependent rectangular function
\begin{equation}
  \varepsilon_k(t) = \varepsilon_{k,0} + \Delta\varepsilon_k \cdot \text{rect}
	\left( \frac{t \text{ mod } T_\text{rep}}{t_\text{puls}} - \frac{1}{2} \right)
\end{equation}
with ${\Delta \varepsilon_k = -(1 - J_0) \varepsilon_{k,0}}$,
where the rectangular function is defined as
\begin{equation}
  \text{rect}(t) = \begin{cases} 0, & \lvert t \rvert > \frac{1}{2} \\ 1, & \lvert t \rvert <\frac{1}{2} \end{cases} .
\end{equation}
This represents the rectangular pulse as shown above in Fig.~\ref{fig:rectangle}.

If we take the limit ${t_\text{puls} \to 0}$ while keeping the area of the pulse ${(1-J_0)t_\text{puls}}$ constant,
the prefactor ${\Delta\varepsilon_k}$ of the rectangular function diverges to $\infty$.
Then, this term dominates in the equations of motion.
Neglecting all non-diverging terms in the equations,
we arrive at equations that describe a simple rotation about the $z$ axis
with angular frequency ${2 \Delta\varepsilon_k = 2(J_0 - 1)\varepsilon_{k, 0}}$ having set ${\hbar=1}$.
Then, the effect of the pulse is fully captured by the rotation matrix
\begin{subequations}
\begin{equation}
  \vec{s}_k (t_\text{puls}) =  \begin{pmatrix}
                                    \cos(\varphi_k) & -\sin(\varphi_k) & 0 \\
                                    \sin(\varphi_k) & \cos(\varphi_k) & 0 \\
                                    0 & 0 & 1
                                \end{pmatrix} \vec{s}_k(0)
  \label{eq:drehung}
\end{equation}
with the rotation angle
\begin{align}
  \varphi_k &= 2 \Delta\varepsilon_k t_\text{puls} \\
            &= 2(J_0 - 1) \varepsilon_{k, 0} t_\text{puls}
            \eqqcolon -2\varphi \varepsilon_{k, 0} .
\end{align}
\end{subequations}
The parameters $J_0$ and $t_\text{puls}$ are combined to the parameter ${\varphi = (1 - J_0) t_\text{puls}}$
which we keep constant in the idealization by keeping the rectangular area constant.
In this way, we reduce the dynamics during the pulse to an instantaneous rotation around the $z$ axis.
Solving the whole dynamics means that we alternate between the original equations of motion~\eqref{eq:dgl}
\emph{between} the pulses and the instantaneous rotation~\eqref{eq:drehung}.
This corresponds to the pulse shape of a $\delta$, significantly simplifies
the numerics, and reduces the number of parameters.

\begin{figure}[htb]
    \centering
    \includegraphics[width=\columnwidth]{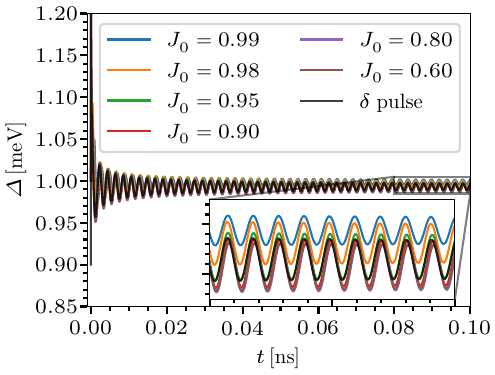}
    \caption{Comparison of the dynamics for different rectangular pulses of the same area
    ${(1-J_0)t_\text{puls} = \SI{0.01}{ps}}$ and the corresponding $\delta$ pulse.
    The convergence of the results for the rectangular pulses with decreasing duration to the $\delta$ pulse is good.}
    \label{fig:to_delta_pulse_dynamics1}
\end{figure}

\begin{figure}[htb]
    \centering
    \includegraphics[width=\columnwidth]{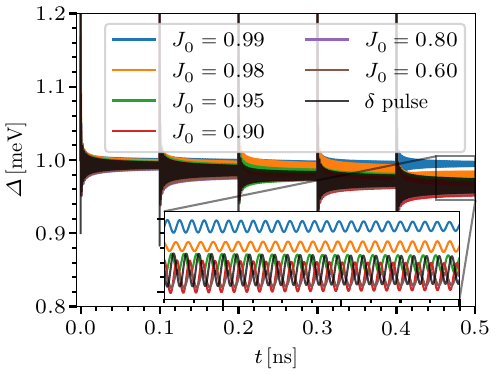}
    \caption{Comparison of the dynamics for trains of different rectangular pulses
		of the same area ${(1-J_0)t_\text{puls} = \SI{0.01}{ps}}$
          and the corresponding train of $\delta$ pulses.
          Five pulses with a repetition time of ${T_\text{rep} = \SI{100}{ps}}$ are shown.}
    \label{fig:to_delta_pulse_dynamics}
\end{figure}

First, we compare the results of rectangular pulses and the $\delta$ pulse
to investigate if the idea to approximate the rectangular pulse by an instantaneous one makes sense.
In Figs.~\ref{fig:to_delta_pulse_dynamics1} and~\ref{fig:to_delta_pulse_dynamics},
we compare the dynamics of the order parameter for several rectangular pulses of the same area
to the dynamics of a $\delta$ pulse with the same area.
The agreement between the shortest rectangular pulse and the $\delta$ pulse is good enough
to conclude that the rectangular pulse converges reasonably well to the $\delta$ pulse
in the limit ${t_\text{puls} \to 0}$.
This can also be confirmed by a comparison of the distributions of the pseudospin
expectation values after the first pulse, i.e., at  ${t = T_\text{rep} = \SI{100}{ps}}$.

\begin{figure}[htb]
    \centering
    \includegraphics[width=\columnwidth]{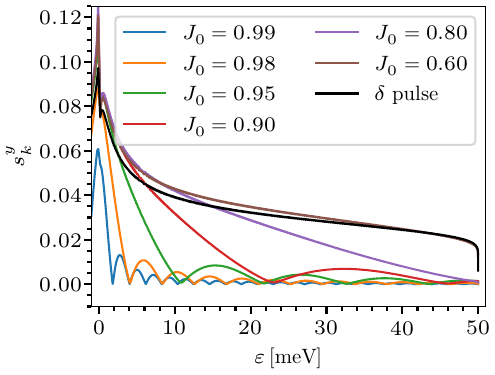}
    \caption{Comparison of the envelopes of the pseudospin distribution
		after a delay time ${t_\text{del} = \SI{100}{ps}}$ after a single pulse.}
    \label{fig:to_delta_pulse_sy1}
\end{figure}

Figure~\ref{fig:to_delta_pulse_sy1} shows the envelope of the $y$ component of the pseudospin
expectation values after a single pulse, but with a certain delay of ${t_\text{del} = \SI{100}{ps}}$.
As was to be expected the nodes of the envelope vanish for the $\delta$ pulse because we established
an inverse relationship between the distance between the nodes and the pulse duration.
Hence the distance tends to infinity for vanishing pulse duration.
The convergence to the result induced by the $\delta$ pulse is satisfactory, but far
from perfect which has to be kept in mind for the subsequent investigations.

\begin{figure}[htb]
    \centering
    \includegraphics[width=\columnwidth]{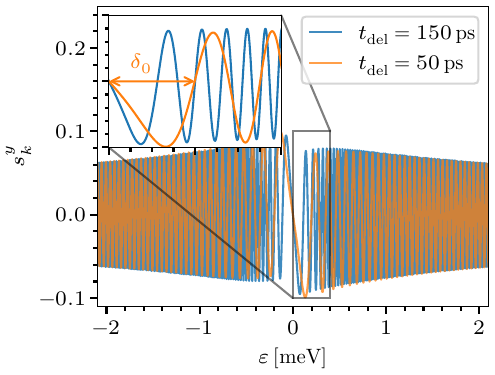}
    \caption{Zoom of the pseudospin distribution at different delay times after	a pulse.}
    \label{fig:delta_pulse_sy1_small}
\end{figure}

\begin{figure}[htb]
    \centering
    \includegraphics[width=\columnwidth]{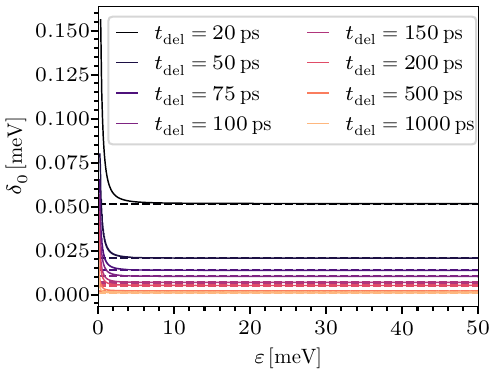}
    \caption{Analysis of the oscillation period of the pseudospin values as function of
		the delay time $t_\text{del}$ after a pulse.
            The distances between the zeros are plotted vs.\ the energy $\varepsilon$.
            The dashed lines indicate the converged limit for high energies $\varepsilon$.}
    \label{fig:zerocrossings_energy_phi0.020}
\end{figure}

\begin{figure}[htb]
    \centering
    \includegraphics[width=\columnwidth]{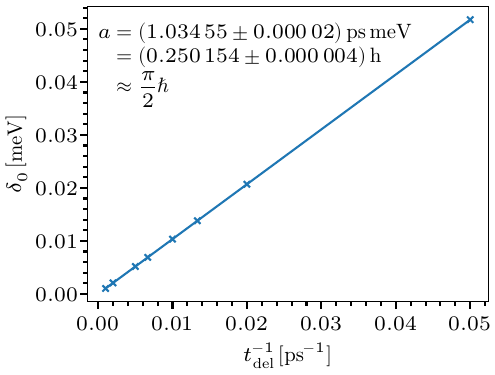}
    \caption{The limits shown by dashed lines in Fig.~\ref{fig:zerocrossings_energy_phi0.020}
		are compared to the inverse delay time $t_\text{del}^{-1}$.
    The linear fit works perfectly.}
    \label{fig:zerocrossings_phi0.020}
\end{figure}

In order to study the effect of the time between the pulses quantitatively, we first focus
on a single pulse and compute the pseudospin values after some delay time $t_\text{del}$
after this pulse in Fig.~\ref{fig:delta_pulse_sy1_small}.
It can be seen that the envelope does hardly change while the oscillations in the fine structure
are strongly modified.
To analyze these modifications, we calculate the distances between neighboring zeros
in the fine structure and plot them in Fig.~\ref{fig:zerocrossings_energy_phi0.020}
vs.\ the energy $\varepsilon$.
Close to ${\varepsilon = 0}$, the node distances are quite large,
but converge quickly to a constant value for increasing $\varepsilon$.
We display the distances $\delta_0$ in the converged regime at large $\varepsilon$
against the inverse delay time ${1/t_\text{del}}$ in Fig.~\ref{fig:zerocrossings_phi0.020}.
This unambiguously shows that the converged distances are given by ${c/t_\text{del}}$
with the prefactor ${a = \frac{\pi}{2}\hbar}$; recall that we set $\hbar$ to one in the calculations.
Thus, the energy interval of a full oscillation is given by ${\hbar \pi/t_\text{del}}$.

In App.~\ref{sec:A1}, we show that the $y$ component at a time delay $t\dl$ after a pulse is approximated by
\begin{equation}
  s_{k,1}^y(t\dl)
    = - 2 \frac{\Delta_0}{E_{k,0}} \varphi \varepsilon_{k,0} \cos\left( 2E_{k,0} t\dl \right) .
\end{equation}
Thus, for a given delay time $t\dl$, the pseudospin distribution is proportional to
\begin{subequations}
\begin{align}
    \cos\left(2E_{k,0} t\dl\right)
    &= \cos \left( 2 t\dl \sqrt{\varepsilon_k^2 + \Delta_0^2} \right) \\
    &\approx \cos \left( 2 t\dl \varepsilon_k\right) \text{ for } \varepsilon_k \gg \Delta_0  ,
\end{align}
\end{subequations}
i.e., to cosine.
As we are investigating the pseudospin distribution vs.\ $\varepsilon_k$,
the prefactor of $\varepsilon_k$ in the argument of the cosine is a time given by ${2t\dl}$.
Hence, the distance between two zeros takes the value
\begin{equation}
  \delta_0 =  \frac{\pi}{2} t\dl^{-1}  .
\end{equation}
The same factor occurs in the approximation \eqref{eq:sinc_analytical} for the rectangular pulse,
which again corroborates that the $\delta$ pulse induces the same fine structure
as the rectangular pulse except for the sinc envelope.


\section{Conclusion}\label{sec:Conclusion}

The present study was motivated by the unconventional structures~\cite{jaschke2017,kleinjohann2018,schering2020} and the reduction of entropy~\cite{uhrig2020}
that can be achieved by periodic trains of optical pulses applied to the electron spin in quantum dots
with hyperfine interaction to their baths of nuclear spins.
These systems are conveniently described by central spin models.
In the Anderson pseudospin formalism the time-dependent mean-field theory for BCS superconductors
bears strong similarity with the central spin model used for the spins in quantum dots.
Hence, a study of the effects of optical pulses and trains of them  on superconductors was called for
and represents the main goal of the present study.
Indeed, remarkable structures in the distribution of modes appeared.

We numerically investigated the BCS model under the influence of optical pulses.
Their electromagnetic field couples to the superconducting electrons through Peierls substitution.
For optical pulses the effect of the fast oscillating fields is to reduce the bare dispersion
upon averaging over these oscillations.
We found non-equilibrium structures in the occupation of the electronic modes
stemming from such perturbations of the superconductors.
The structures depend on the amplitude and duration of the pulses.
We found a detailed rich distribution of the expectation values of the pseudospins of which the envelope is governed by
the Fourier transform of the pulse shape and the fast oscillations by the delay time elapsed after the pulse.
This is confirmed by an analytical approximation based on the linearization of the equations of motion.
In addition, we found a periodic dependence of the energy deposited in the system on the duration of the pulses.
The periodicity can be explained by the coherent precession of the pseudospins such that
the overall effect depends roughly on whether half revolutions fit into the pulse or not, i.e.,
the angular frequency of the periodicity is given by $2\Delta_0$ where $\Delta_0$ is the initial
superconducting gap.
The factor of 2 appears in the limit of very weak pulses only.

The fascinating structures that our theoretical simulations unearthed call for experimental verification.
It would be highly interesting to investigate how such structures appear in experiments,
i.e., in conventional superconductors which are pumped by ultrashort optical pulses.
The results presented here provide a guideline to such experiments by showing what is to be expected.
For realization, of course, one must be aware of the effects of heating.
These do not appear in our treatment where we average the effect of the Peierls phases on the hopping.
We stress that our setup does not require any particular value of the optical frequency
as long as ${f_\text{opt} t_\text{puls}\gg 1}$ and ${hf_\text{opt}/\Delta_0\gg 1}$  holds.
The material system best suited for realization should have a transparency window,
i.e., a window of optical frequencies, within which no or hardly any resonance occurs so that
no absorption takes place.
Certainly, this represents quite a challenge to find and to create.

On the theoretical side, the next steps should include explicit calculations of further observables
that can be directly measured in an experiment, for instance, the optical conductivity.
Another interesting extension consists in the inclusion of dissipation
because such effects are surely present in experiments.
They can be included either phenomenologically as in Ref.~\onlinecite{cui2019a}
in theoretical treatments or by either to higher order diagrams inducing finite imaginary parts
in the self-energies or by resorting to Lindbladian dynamics for the pseudospins.
These intriguing issues are beyond the scope of the present investigation and thus are left to future research.

\begin{acknowledgments}
This research was supported by the Stiftung Mercator in project no. KO-2021-0027.
We acknowledge very helpful discussions with J.~Alth\"user, A.~B\"ohmer, J.~B\"unemann, I.~Eremin, and J.~Stolze.
\end{acknowledgments}


\appendix

\section{Analytical approximation for the \texorpdfstring{$y$}{y} component of the pseudospins}
\label{sec:A1}

We start from the equations of motion
\begin{subequations}
  \begin{align}
    \dot{s}_k^x &= -2\varepsilon_k s_k^y \\
    \dot{s}_k^y&= 2 \varepsilon_k s_k^x + VJ_x s_k^z \\
    \dot{s}_k^z &= -VJ_x s_k^y
  \end{align}
\end{subequations}
where $ J_x = \frac{1}{N}\sum_k s_k^x $.
In general, the effect of the pulse on the dispersion can be written as
\begin{equation}
    \varepsilon_k(t) = \varepsilon_{k,0} + \lambda f_k(t)
\end{equation}
with a small parameter $\lambda$ and a time-dependent function $f_k(t)$, for instance
of rectangular shape in our study.
As long as the pulse depth is small, the dynamics are occuring as a small perturbation of the equilibrium solution in linear response
\begin{subequations}
\begin{equation}
  s_k^\alpha = s_{k,0}^\alpha + \lambda s_{k,1}^\alpha \quad \text{with} \quad \alpha\in\{x,y,z\}
\end{equation}
and
\begin{equation}
  J_x = J_{x,0} + \lambda J_{x,1}  .
\end{equation}
\end{subequations}
Inserting this into the equations of motion leads to
\begin{subequations}
  \begin{align}
    \begin{split}
    \dot{s}_{k,0}^x + \lambda\dot{s}_{k,1}^x &= -2\varepsilon_{k,0}s_{k,0}^y \\
                                    &\phantom{=}+ \lambda (-2\varepsilon_{k,0}s_{k,1}^y - 2f_k(t)s_{k,0}^y) \\
                                    &\phantom{=}- \lambda^2 2f_k(t)s_{k,1}^y
    \end{split} \\[2ex]
    \begin{split}
    \dot{s}_{k,0}^y + \lambda\dot{s}_{k,1}^y &= 2\varepsilon_{k,0}s_{k,0}^x + gJ_{x,0}s_{k,0}^z \\
                                    &\phantom{=}+ \lambda (2\varepsilon_{k,0}s_{k,1}^x + 2f_k(t)s_{k,0}^x \\
                                    &\phantom{= + \lambda (}+ VJ_{x,0}s_{k,1}^z + VJ_{x,1}s_{k,0}^z) \\
                                    &\phantom{=}+ \lambda^2 (2f_k(t)s_{k,1}^x + VJ_{x,1}s_{k,1}^z)
    \end{split} \\[2ex]
    \begin{split}
    \dot{s}_{k,0}^z + \lambda\dot{s}_{k,1}^z &= -gJ_{x,0}s_{k,0}^y \\
                                    &\phantom{=}- \lambda (VJ_{x,0} s_{k,1}^y + VJ_{x,1}s_{k,0}^y) \\
                                    &\phantom{=}- \lambda^2 VJ_{x,1}s_{k,1}^y  .
    \end{split}
  \end{align}
\end{subequations}
Next, we neglect all terms of order $\lambda^2$ and split the equations into two parts, namely of zeroth and of first order.
The zeroth-order equations are given by
\begin{subequations}
    \begin{align}
        \dot{s}_{k,0}^x &= -2\varepsilon_{k,0} s_{k,0}^y \\
        \dot{s}_{k,0}^y &= 2 \varepsilon_{k,0} s_{k,0}^x + VJ_{x,0} s_{k,0}^z \\
        \dot{s}_{k,0}^z &= -VJ_{x,0} s_{k,0}^y
    \end{align}
\end{subequations}
and the first-order equations are given by
\begin{subequations}
    \begin{align}
        \dot{s}_{k,1}^x &= -2\varepsilon_{k,0}s_{k,1}^y -2f_k(t)s_{k,0}^y \\
        \begin{split}
        \dot{s}_{k,1}^y &= 2\varepsilon_{k,0}s_{k,1}^x + VJ_{x,0}s_{k,1}^z \\
                        &\phantom{=} + 2f_k(t) s_{k,0}^x + VJ_{x,1}s_{k,0}^z
        \end{split} \\
        \dot{s}_{k,1}^z &= -VJ_{x,0} s_{k,1}^y - VJ_{x,1}s_{k,0}^y  .
    \end{align}
\end{subequations}
The zeroth order reproduces the equilibrium solution which we insert into the first-order equations.
With the matrix
\begin{equation}
    \mat{M}\coloneqq
    \begin{pmatrix}
        0 & -2\varepsilon_{k,0} & 0 \\
        2\varepsilon_{k,0} & 0 & VJ_{x,0} \\
        0 & -VJ_{x,0} & 0
    \end{pmatrix}
\end{equation}
the resulting equations can be written as
\begin{subequations}
\begin{equation}
    \dot{\vec{s}}_{k,0}(t) = \mat{M} \vec{s}_{k,0}(t)
\end{equation}
and
\begin{equation}
  \dot{\vec{s}}_{k,1}(t) = D_1 + D_2 + D_\text{ext}
  \label{eqn:dgl}
\end{equation}
with
\begin{align}
  D_1 &= \mat{M} \vec{s}_{k,1}(t) \label{eq:D_1}\\
  D_2 &= VJ_{x,1}(t) \begin{pmatrix} 0 \\ s_{k,0}^z(t) \\ -s_{k,0}^y(t) \end{pmatrix} \label{eq:D_2} \\
  D_\text{ext} &= 2f_k(t) \begin{pmatrix} -s_{k,0}^y(t) \\ s_{k,0}^x(t) \\ 0 \end{pmatrix}  .
	\label{eq:D_ext}
\end{align}
\end{subequations}
Of course, the equilibrium solution is known
\begin{equation}
  s_{k,0}^x = \frac{\Delta_0}{E_{k,0}} ,\quad s_{k,0}^y = 0 ,\quad s_{k,0}^z = -\frac{\varepsilon_{k,0}}{E_{k,0}}
\end{equation}
with $E_{k,0} = \sqrt{\varepsilon_{k,0}^2 + \lvert\Delta_0\rvert^2}$.

\begin{figure}[htb]
  \centering
  \includegraphics[width=\columnwidth]{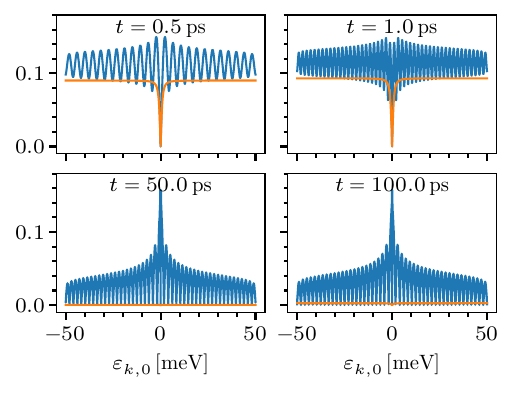}
  \caption{Measures of ${\vert D_1\vert}$ (blue lines) and ${\vert D_2\vert}$ (orange lines) as defined in
	Eqs.~\eqref{eq:D_1} and~\eqref{eq:D_2} for a rectangular pulse with
    ${J_0 = \num{0.99}}$, ${t_\text{puls} = \SI{1}{ps}}$ at various delay times $t_\text{del}$.}
  \label{fig:compare_summands_rect}
\end{figure}
\begin{figure}[htb]
  \centering
  \includegraphics[width=\columnwidth]{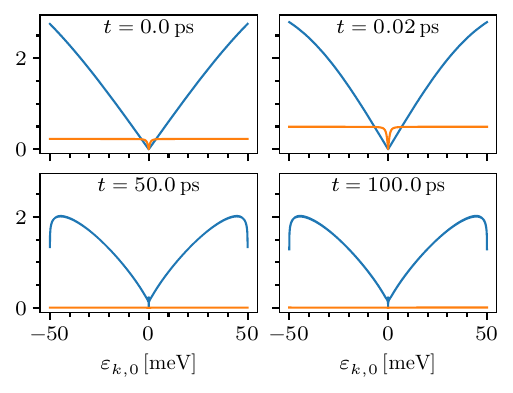}
  \caption{Same as in Fig.~\ref{fig:compare_summands_rect} for a $\delta$ pulse with
    ${\varphi = (J_0 - 1) t_\text{puls} = \SI{0.01}{ps}}$ at various delay times ${t\dl}$.}
  \label{fig:compare_summands_delta}
\end{figure}

We want to neglect $D_2$ in Eq.~\eqref{eqn:dgl} to reach a simpler equation that can be solved analytically.
To verify that this is a valid approximation, we compare $D_2$ to $D_1$.
For this purpose, we need a measure for the size of $D_1$,
namely the projections of the spin configuration onto the eigenvectors of $\mat{M}$, namely
${\lvert \lambda_{2/3} \rvert \sqrt{\lvert\braket{\vec{s}_{k,1} \vert \vec{e}_2 }\rvert^2
+ \lvert\braket{\vec{s}_{k,1} \vert \vec{e}_3 }\rvert^2}}$.
Given the equilibrium solution, the absolute value of $D_2$ reads
${V \left\lvert J_{x,1} \frac{\varepsilon_{k,0}}{E_{k,0}} \right\rvert}$.
To compare the two terms, we plot them against $\varepsilon_{k,0}$ for exemplary pulse parameters and various times
in Fig.~\ref{fig:compare_summands_rect} for the rectangular pulse and
in Fig.~\ref{fig:compare_summands_delta} for the $\delta$ pulse.
Values for $\vec{s}_{k,1}$ are taken from the numerical evaluation of the differential equations.

For both kinds of pulses, the approximation is valid for large values of ${\vert\varepsilon_{k,0}\vert}$
because $D_2$ is significantly smaller than $D_1$.
For the $\delta$ pulse (Fig.~\ref{fig:compare_summands_delta})
$D_2$ is almost everywhere smaller than $D_1$ except in the immediate vicinity of ${\varepsilon_{k,0}=0}$.
After the rectangular pulse (lower row in Fig.~\ref{fig:compare_summands_rect}),
$D_1$ also dominates clearly over $D_2$.
But during the rectangular pulse (upper row in Fig.~\ref{fig:compare_summands_rect}),
$D_2$ is of the same order of magnitude as $D_1$ up to moderate values of the argument.
We conclude that the approximation is justified except for small values of  ${|\varepsilon_{k,0}|}$.
This is in line with our previous comparison where the analytic approximations worked well
for large ${|\varepsilon_{k,0}|}$, but showed deviations for small values.


Neglecting the term $D_2$, the differential equation can be solved analytically by multi-dimensional
variation of constants
\begin{subequations}
\begin{align}
    \vec{s}_{k,1}(t) &= \exp(\mat{M}t) \int_0^t \exp(-\mat{M}t')\vec{v}(t')\dx{t'}
    \intertext{with}
    \vec{v}(t) &= 2f_k(t) \begin{pmatrix} -s_{k,0}^y(t) \\ s_{k,0}^x(t) \\ 0 \end{pmatrix} .
\end{align}
\end{subequations}
The eigenvalues and eigenvectors of ${-\mat{M}}$ are given by
\begin{subequations}
\begin{align}
    \lambda_1 &= 0 &
    \vec{e}_1 &= \frac{1}{2E_{k,0}}\begin{pmatrix} -VJ_{x,0} \\ 0 \\ 2\varepsilon_{k,0} \end{pmatrix} \\
    \lambda_2 &= -2iE_{k,0} &
    \vec{e}_2 &= \frac{1}{2\sqrt{2}\,E_{k,0}} \begin{pmatrix} 2\varepsilon_{k,0} \\ -2iE_{k,0} \\ VJ_{x,0} \end{pmatrix} \\
    \lambda_3 &= 2iE_{k,0} &
    \vec{e}_3 &= \frac{1}{2\sqrt{2}\,E_{k,0}} \begin{pmatrix} 2\varepsilon_{k,0} \\ 2iE_{k,0} \\ VJ_{x,0} \end{pmatrix}
\end{align}
with
\begin{align}
  2 E_{k,0}
    &= \sqrt{(2\varepsilon_{k,0})^2 + \left({V}J_{x,0}/N\right)^2} \\
    &= \sqrt{(2\varepsilon_{k,0})^2 + (2\Delta_0)^2}  .
\end{align}
\end{subequations}

Then, the exponential function of $\mat{M}$ and its inverse are given by
\begin{subequations}
\begin{equation}
    \exp(\mp\mat{M}t') = \mat{U} \exp(\pm\mat{D}t') \mat{U}^{-1}
\end{equation}
with the unitary matrix
\begin{align}
    \mat{U}
    &= \begin{pmatrix} \vert & \vert & \vert \\ \vec{e}_1 & \vec{e}_2 & \vec{e}_3 \\ \vert & \vert & \vert \end{pmatrix} \\
      \exp(\pm\mat{D}t')
    &= \begin{pmatrix} e^{\pm\lambda_1 t'} & 0 & 0 \\ 0 & e^{\pm\lambda_2 t'} & 0 \\ 0 & 0 & e^{\pm\lambda_3 t'} \end{pmatrix}  .
\end{align}
\end{subequations}
This enables us to write the complete solution as
\begin{subequations}
\begin{align}
    \vec{s}_{k,1}(t)
        &= e^{\mat{M}t} \int_0^t e^{-\mat{M}t'} \vec{v}(t') \dx{t'} \\
        &= \mat{U} e^{-\mat{D}t} \int_0^t e^{\mat{D}t'} \mat{U}^{-1} \vec{v}(t') \dx{t'}  .
\end{align}
\end{subequations}
From these formulae, we deduce the $y$ component obtaining
\begin{align}
  \begin{split}
    s_{k,1}^y(t)
    &= e_1^y\,e^{-\lambda_1 t} \int_0^t e^{\lambda_1 t'} \vec{e}_1^* \cdot \vec{v}(t') \dx{t'} \\
      &\quad + e_2^y\,e^{-\lambda_2 t} \int_0^t e^{\lambda_2 t'} \vec{e}_2^* \cdot \vec{v}(t') \dx{t'} \\
      &\quad + e_3^y\,e^{-\lambda_3 t} \int_0^t e^{\lambda_3 t'} \vec{e}_3^* \cdot \vec{v}(t') \dx{t'}  .
    \end{split}
\end{align}
The last two terms are the complex conjugate of each other because
${\vec{e}_2 = \vec{e}_3^*}$ and ${\lambda_2 = \lambda_3^*}$.
Thus, one obtains
\begin{align}
    s_{k,1}^y(t)
    &= 2\Re\Big(-\frac{i}{2E_{k,0}} e^{2iE_{k,0} t} \int_0^t e^{-2iE_{k,0} t'} f_k(t') \cdot \nonumber \\
    & \quad \left(-2\varepsilon_{k,0} s_{k,0}^y (t') + 2iE_{k,0} s_{k,0}^x (t')\right) \dx{t'}\Big) .
\end{align}
Inserting the equilibrium solution, we arrive at
\begin{align}
    s_{k,1}^y(t)
    &= 2\Re\left(e^{2iE_{k,0} t} \int_0^t e^{-2iE_{k,0} t'} f_k(t') \frac{\Delta_0}{E_{k,0}} \dx{t'}\right) .
\end{align}
For ${t>t_\text{puls}}$, the integral can be extended
to ${\pm\infty}$ because ${f_k(t') = 0}$ outside of the support of the pulse so that we obtain
\begin{align}
    s_{k,1}^y(t)
    &= 2\Re\left(e^{2iE_{k,0} t} \int_{-\infty}^\infty e^{-2iE_{k,0} t'} f_k(t') \frac{\Delta_0}{E_{k,0}} \dx{t'}\right) .
\end{align}
Next, we shift the time variable by ${t_\text{puls}/2}$ because ${f_k(t')}$ is symmetric around ${t'=t_\text{puls}/2}$
\begin{align}
  \begin{split}
    s_{k,1}^y(t) = 2\frac{\Delta_0}{E_{k,0}} \Re\Bigg(
      &e^{2iE_{k,0} \left(t - {t_\text{puls}/2}\right)} \\
      &\int_{-\infty}^\infty e^{-2iE_{k,0} t'} \tilde{f}_k(t') \dx{t'}\Bigg) .
  \end{split}
\end{align}
The shifted function ${\tilde{f}_k(t')}$ is a real and even function so that its
Fourier integral is real yielding
\begin{subequations}
\begin{align}
    s_{k,1}^y(t)
    &= 2\frac{\Delta_0}{E_{k,0}}\Re\left( e^{2iE_{k,0} \left(t - {t_\text{puls}/2}\right)}\right) \notag \\
      &\quad\cdot \int_{-\infty}^\infty e^{-2iE_{k,0} t'} \tilde{f}_k(t') \dx{t'} \\
    &= 2 \frac{\Delta_0}{E_{k,0}} \cos\left( 2E_{k,0} \left(t - {t_\text{puls}/2}\right) \right)
		\mathcal{F}(\tilde{f}_k(t'))  .
    \label{eqn:approx}
\end{align}
\end{subequations}
with ${\mathcal{F}(\tilde{f}_k(t'))}$ being the Fourier transform of ${\tilde{f}_k(t')}$.
This provides us with an analytical approximation for the $y$ component of the spin dynamics for small pulse depths.
One can insert various pulse shapes and compare the results to the numerical solutions.


\subsection{Rectangular pulse}
\begin{figure}[htb]
  \centering
  \includegraphics[width=\columnwidth]{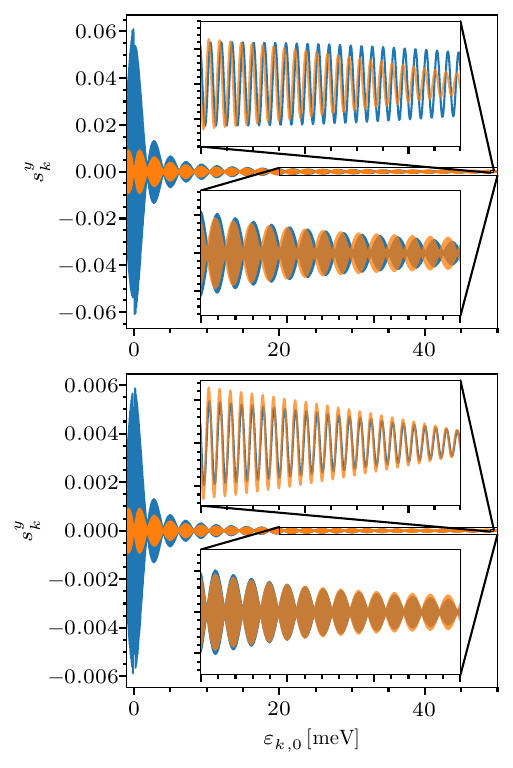}
  \caption{Comparison of the analytical approximation (orange lines) and the numerical result (blue lines)
  for a rectangular pulse of duration ${t_\text{puls} = \SI{1}{ps}}$
  at two different pulse depths of ${J_0 = \num{0.99}}$ (upper panel) and ${J_0 = \num{0.999}}$ (lower panel).
  We compare the $y$ component of the pseudospin distribution after a single pulse and a delay time of
  ${t_\text{del} = \SI{100}{ps}}$.}
  \label{fig:comparison_rectangle}
\end{figure}

The shifted rectangular pulse is given by
\begin{equation}
    \tilde{f}_k(t') = (J_0 - 1) \varepsilon_{k,0}
        \begin{cases}
            1 & \lvert t \rvert \leq {t_\text{puls}/2} \\
            0 & \lvert t \rvert > {t_\text{puls}/2}
        \end{cases}
\end{equation}
and its Fourier transform reads
\begin{subequations}
\begin{align}
    \mathcal{F}(\tilde{f}_k(t'))
    &= \int_{-\infty}^\infty e^{-2iE_{k,0} t'} \tilde{f}_k(t') \dx{t'} \\
    &= (J_0 - 1) \varepsilon_{k,0} t_\text{puls} \,\mathrm{sinc}\left( E_{k,0} t_\text{puls} \right)
\end{align}
\end{subequations}
with the sinc function $\mathrm{sinc}(x) = \sin(x)/x$.
Thus, the $y$ component of the spin dynamics according to Eq.~\ref{eqn:approx} is given by
\begin{align}
  s_{k,1}^y(t)
    &= 2 \frac{\Delta_0}{E_{k,0}} (J_0 - 1) \varepsilon_{k,0} t_\text{puls}
		\cos\left( 2E_{k,0} \left(t - {t_\text{puls}/2}\right) \right)
      \nonumber \\
     &\hphantom{=}\cdot \mathrm{sinc}\left( E_{k,0} t_\text{puls} \right) .
\end{align}

This analytical result is compared to the numerical solution in Fig.~\ref{fig:comparison_rectangle} for two pulse depths.
As expected, the approximation does not work for small values of $\varepsilon_{k,0}$,
but the agreement for higher values is good.
For ${J_0 = \num{0.99}}$, the results are slightly shifted,
but the overall shape of approximate and numerical behavior is the same.
The envelope of the numerical results is captured very nicely by the $\mathrm{sinc}$ function.
The fast oscillations are reproduced by the cosine function very well.


\subsection{\texorpdfstring{$\delta$ pulse}{𝛿 pulse}}

The $\delta$ pulse
\begin{equation}
    \tilde{f}_k(t') = -\varphi \varepsilon_{k,0} \delta(t')
\end{equation}
implies the Fourier transform
\begin{subequations}
\begin{align}
    \mathcal{F}(\tilde{f}_k(t'))
    &= \int_{-\infty}^\infty e^{-2iE_{k,0} t'} \tilde{f}_k(t') \dx{t'} \\
    &= -\varphi \varepsilon_{k,0} ,
\end{align}
\end{subequations}
which leads to
\begin{equation}
  s_{k,1}^y(t)
    = - 2 \frac{\Delta_0}{E_{k,0}} \varphi \varepsilon_{k,0} \cos\left( 2E_{k,0} t \right) .
\end{equation}

\begin{figure}[htb]
    \centering
    \includegraphics[width=\columnwidth]{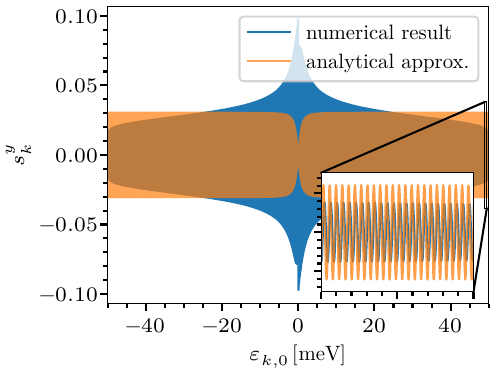}
    \caption{Comparison  of the analytical approximation and the numerical result for a $\delta$ pulse
		at ${t\dl = \SI{100}{ps}}$ for amplitude ${\varphi = \SI{0.01}{ps}}$ corresponding  to a rectangular pulse of duration
		${t_\text{puls} = \SI{1}{ps}}$ and pulse depth ${1-J_0 = \num{0.01}}$.}
    \label{fig:comparison_delta}
\end{figure}

We compare this approximation to the numerical result in Fig.~\ref{fig:comparison_delta}.
Here again, the fast oscillations are captured well by the cosine function,
but the envelope deviates from the numerical result.
For both pulse shapes, there seems to be some additional envelope function missing.
This might be caused by neglecting $D_2$ in \eqref{eqn:dgl}.
In addition, we recall that the employed approximation is valid for small pulse depths
which is not the appropriate regime for the application to a $\delta$ pulse.

\bibliography{bibliography}

\end{document}